\documentclass[aps,prl,superscriptaddress,showpacs,noshowkeys,amsmath,amssymb,amsfonts,reprint]{revtex4-1}
\usepackage{amsfonts}
\usepackage{graphicx}
\usepackage{amsmath}
\usepackage{times}
\usepackage{amssymb}
\usepackage{color}
\usepackage{bm}
\usepackage{bbm}
\usepackage{amsfonts, amsmath, amsthm, amssymb} 
\usepackage{array}
\usepackage[colorlinks,bookmarks=false,citecolor=blue,linkcolor=red,urlcolor=blue]{hyperref}
\usepackage{comment}

\newcommand{\hide}[1]{}
\newcommand{\be}{\begin{equation}}
\newcommand{\bee}{\begin{equation*}}
\newcommand{\ee}{\end{equation}}
\newcommand{\eee}{\end{equation*}}
\newcommand{\bearre}{\begin{eqnarray*}}
\newcommand{\eearre}{\end{eqnarray*}}
\newcommand{\bearr}{\begin{eqnarray}}
\newcommand{\eearr}{\end{eqnarray}}

\begin{document}
\title{Competing orders for the colloidal kagome ice:  the importance of in-trap motion of the particles}
\author{Anne Le Cunuder}
\email{anne.le-cunuder@fmc.ub.edu}
\affiliation{
Departament de F\'isica de la Mat\`eria Condensada, Universitat de Barcelona, Barcelona, Spain}
\author{Ir\'en\'ee Fr\'erot}
\email{irenee.frerot@icfo.eu}
\affiliation{
ICFO - Institut de Ciencies Fotoniques, The Barcelona Institute of Science and Technology, Av. Carl Friedrich Gauss 3, 08860 Castelldefels (Barcelona), Spain}
\author{Antonio Ortiz-Ambriz}
\affiliation{
Departament de F\'isica de la Mat\`eria Condensada, Universitat de Barcelona, Barcelona, Spain}
\affiliation{Institut de Nanoci\`{e}ncia i Nanotecnologia, Universitat de Barcelona,  Barcelona, Spain.}
\author{Pietro Tierno}
\email{ptierno@ub.edu}
\affiliation{
Departament de F\'isica de la Mat\`eria Condensada, Universitat de Barcelona, Barcelona, Spain}
\affiliation{Institut de Nanoci\`{e}ncia i Nanotecnologia, Universitat de Barcelona,  Barcelona, Spain.}
\affiliation{Universitat de Barcelona Institute of Complex Systems (UBICS), Universitat de Barcelona, Barcelona, Spain.}
\date{\today}

\begin{abstract}
Artificial ice systems have been designed
to replicate paradigmatic phenomena observed in frustrated spin systems.
Here we present a detailed theoretical analysis based on Monte-Carlo simulations of the low energy phases in an artificial colloidal ice system, a recently introduced ice system where an ensemble of repulsive colloids are two-dimensionally confined by gravity to a lattice of double wells at a one-to-one filling.
Triggered by recent results obtained by Brownian dynamics simulations [A. Lib\'al {\it et al.}, Phys. Rev. Lett. \textbf{120}, 027204 (2018)], we analyze the energetics and the phase transitions that occur in the honeycomb geometry (realizing the analogue
of a spin ice system on a kagome lattice) when decreasing the temperature.
When the particles are restricted to occupy the two minima of the potential well, we recover the same phase diagram as the dipolar spin ice system, with a long range ordered chiral ground state. 
In contrast, when considering the particle motion
and their relaxation within the traps, we observe ferromagnetic ordering at low temperature.
This observation highlights the fundamental role played by the continuous motion of the colloids in artificial ice systems.
\end{abstract}
\maketitle
\textit{Introduction.--}
Artificial spin ice systems (ASIs) are networks of nanoscale ferromagnetic islands~\cite{Wan06} or wires~\cite{Tan06}, arranged such that their mutual interactions generate frustration and lead to exotic spin ordering. 
These systems are designed as two-dimensional analogs of water ice~\cite{Pau35,Lie67} and frustrated magnets~\cite{Ste01,Lac09}, and furthermore, they allow one to directly visualize
the spin configurations, and to design at will the lattice
geometry~\cite{Nis13}. 
Like frustrated magnets, ASIs exhibit magnetic charges and topological defects that can be controlled via an external field~\cite{Mol09,Lad10,Mor11,Men11,Pha11,Kap14,Gil14,Ved16}, a feature that enables a variety of potential  applications in magnetic devices~\cite{Wang962}.

An alternative, soft-matter-based, approach to study magnetic frustration phenomena in 2D lattices, relies on buckled arrangements of colloidal particles under strong confinement~\cite{Han08}.
In contrast to ASIs, these systems allow for the observation
of the dynamics in real time through relatively simple optical
microscopy.
In this context, numerical simulations have predicted that electrostatically repulsive colloidal particles confined in a lattice of bistable potentials could reproduce most of the features observed in ASIs, such as the emergence of ice selection rules~\cite{Lib06,Lib121}.
The analogy between ASIs and artificial colloidal ice has been investigated in depth, showing that, for lattice characterized by a single coordination number, the ice manifold is equivalent in both systems~\cite{Nis14,Lib18}.

Recently, two-dimensional colloidal ice systems have been realized in experiments using repulsive paramagnetic particles confined by gravity in a lattice of lithographic microgrooves~\cite{Ort16,Loe16,Lee18}. 
Due to the relatively large size of the particles, this system displays weak thermal fluctuations, which makes the effect of temperature difficult to observe. Instead, to find the low energy states, the system is  quenched by increasing the interaction energy between the particles via an external field.
A similar procedure was used in Brownian dynamics simulations of colloidal ice on a kagome lattice~\cite{libaletal2018}, a geometry which presents an highly degenerate low energy phase in ASI.
In the colloidal ice it was found that, raising the interaction strength between the particles, the system finally orders in a ferromagnetic state. This observation contrasts with Monte-Carlo simulations of dipolar kagome spin ice which show the formation of a different ordered phase at low temperature, in which the spins on the hexagons define chiral and achiral loops~\cite{Mol09a,Che11}. Experimental evidence of such phases was also reported in ASIs by different groups~\cite{Bri08,Lad12,Zha13,canalsetal2016}.

In this article, we report Monte-Carlo simulations of the low energy states of a kagome colloidal ice.
Here, instead of varying the interaction strength, we tune directly the system temperature and consider both the limiting case of immobile particles (in which the system is fully described by an effective spin Hamiltonian) and the continuous case in which particles can relax locally within the bistable traps. 
In the first (discrete) case, we recover a behavior similar to ASIs, while in the second (continuous) case, we observe that the system undergoes a transition to a ferromagnetic phase.
Our findings therefore reconcile the apparent contradiction between the reported results from Brownian simulations of colloidal ice, and the previous theoretical works on the ASIs. 

\begin{figure}[t]
\begin{center}
\includegraphics[width=\columnwidth,keepaspectratio]{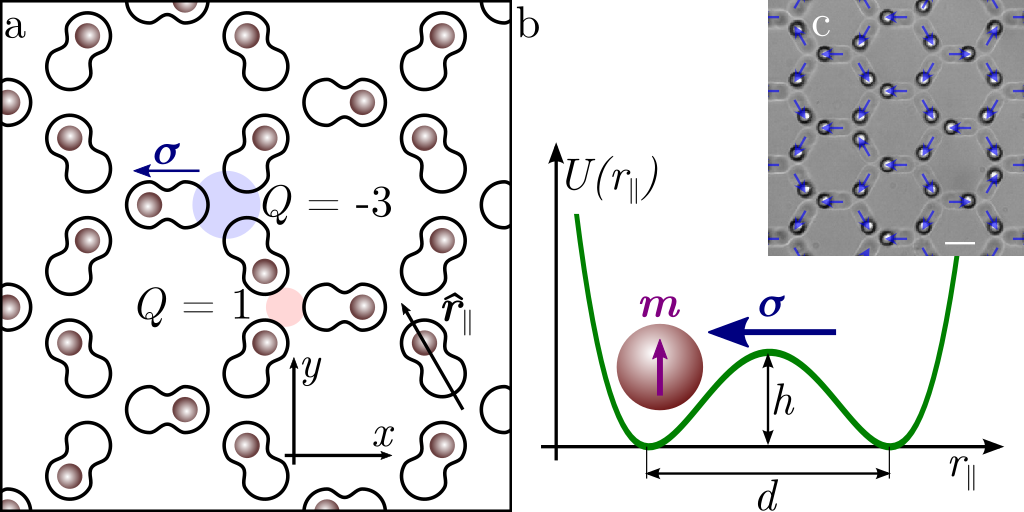} 
\caption{(a) Schematic of the colloidal ice  
which consists in a kagome lattice of double wells filled by repulsive paramagnetic colloids. 
The blue and red disks denote negative ($3$ particles \textit{out}) 
and positive ($2$ particles \textit{in}, $1$ \textit{out}) topological charges $Q$.
(b) Double-well potential characterized by 
a distance $d$ between wells and a trap elevation $h$, 
being 
$\sigma$ the associated spin. 
(c) Image from~\cite{Ort16}
showing an experimental kagome colloidal ice system
(lattice constant $a=44 \rm{\mu m}$)
where the blue arrows denote the direction
of the spins $\sigma$.}
\label{Figure1}
\end{center}
\end{figure}

\textit{Experimental motivation.--}
Figs.~\ref{Figure1} illustrate the main features of the colloidal system. 
Paramagnetic colloids of volume $V$ are confined by gravity to a kagome lattice of bistable traps, having a lattice constant $a$ [Fig.~\ref{Figure1}(a)].
Each trap is occupied by one particle that can be in one of the two wells located at distance $d$, and separated by a central hill of elevation $h$ [Fig.~\ref{Figure1}(b)]. 
A particle in a trap can switch well because
of thermal fluctuations or interactions with neighbors, but not
leave the double well.
Repulsive interactions between the particles are induced through an external magnetic field $\bm{B}$, applied perpendicular to the particle plane. 
The field induces a dipole moment $\bm{m} = V \chi \bm{B}$ in each particle of magnetic susceptibility $\chi$. 
Thus, pairs of particles interact via an isotropic repulsive potential, $E= J / |\bm{r}_{ij}|^3$, with $\bm{r}_{ij} = |\bm{r}_i - \bm{r}_j|$ and $J=\mu_0 m^2 / (4 \pi)$. 
Here, $\bm{r}_i$ is the position of particle $i$ and $\mu_0= 4 \pi \cdot 10^{-7} \rm{H \cdot m}$.  
The system can be mapped to an ASI by associating a pseudo-spin
$\bm{\sigma}$ to each double well, such that it points towards the well occupied by the particle. Thus, 
a spin flip is induced whenever a particle crosses the central hill.
The topological charge $Q$ associated to each vertex is given by the difference between the number of spins pointing towards the vertex (particles \textit{in}) and the spins pointing away from it (particles \textit{out}), Fig.\ref{Figure1}(a). 
On the kagome lattice the ice rules restricts the accessible configurations to vertices with $Q=\pm 1$. 
  
Before illustrating the simulation results, we
show in Fig.3 the energy of the two configurations that are potential candidates as ground states. 
The ”chiral” phase, shown in the inset on the left, is characterized buy a regular pattern of hexagons of
alternating chirality and was observed for ASIs~\cite{willsetal2002,Che11}. The ferromagnetic phase,
inset on the right, has been recently reported in simulations of the kagome colloidal ice \cite{Lib18}. Thus, the figure shows the dependence on system size of the magnetic energy $E$ for $N$ colloidal particles sitting at the bottom of the traps in the two configurations.
We find that the chiral state always has a lower energy than the ferromagnetic state, a result which is robust regardless of the distance $d$ separating the two minima of a double-well.

\textit{Model Hamiltonian.--} 
In our model, we consider the following Hamiltonian for a system of $N$ interacting particles, ignoring kinetic terms
\begin{equation}
{\cal H}_{\rm coll.} = 
	J \sum_{i=1}^N\sum_{j=i+1}^N \frac{1}{\vert {\bm r}_{i}-{\bm r}_{j} \vert^3}+ 
	\sum_{i=1}^N U({\bm r}_i)~,
\label{H_coll}
\end{equation}
where the external potential $U({\bm r}_i)$ defines a regular pattern of double-wells trapping the colloids which, in this case, are centered in the points defined by a kagome lattice with constant $a$.

\begin{figure}[b]
\begin{center}
\includegraphics[width=\columnwidth,keepaspectratio]{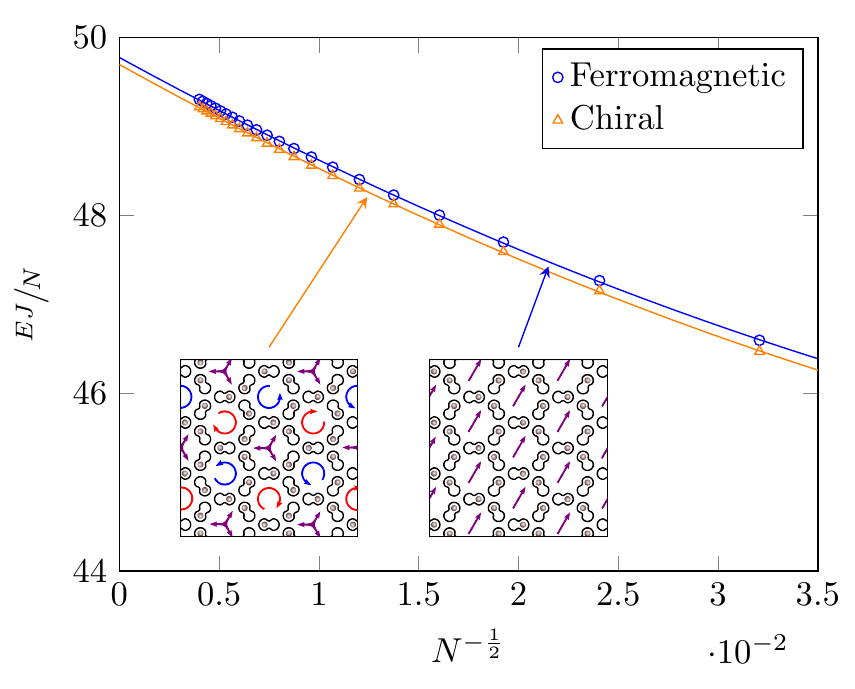}
\caption{Magnetostatic energy 
for $N=3L^2$ interacting particles 
with periodic boundary conditions for 
the chiral (red triangles) and the ferromagnetic (blue circles) phases (up to $L=144$), for $d=0.3$. Inset shows schematics illustrating the two competing orders. Scattered points are numerical calculations, continuous lines are polynomial fits of the form $E/J \sim \alpha N+\beta N^{1/2} +\gamma$, where $\alpha$ is the energy per spin in the thermodynamics limit, while $\beta$ and $\gamma$ are finite-size corrections, fit parameters are in~\cite{Values}.}
\label{fig_GS}
\end{center}
\end{figure}

We perform all simulations with the Hamiltonian defined in Eq.~\eqref{H_coll}. We note that he hierarchy of the phases might be better understood using the formalism described in \cite{nisoli2018}, where a
mapping is proposed by assigning colloids a positive charge,
holes a negative charge, and embedding the system in a fully
occupied system of positive charges.
If we define ${\bm R_i}$ as the center of the double-well which traps particle $i$, and denote with ${\bm \sigma_i}$ 
the associated spin,
we have that
${\bm r}_i = {\bm R}_i + \frac{1}{2}{\bm \sigma}_i$. From the results of \cite{nisoli2018} 
one obtains the following approximate
expression (valid for $\sigma \ll a$):
 \begin{align}
	{\cal H} \approx (3Jd^2 / 4) \sum_{i=1}^N\sum_{j=i+1}^N &\frac{
	    {\bm \sigma}_i \cdot {\bm \sigma}_j - 
	    5({\bm \sigma}_i \cdot \hat{\bm R}_{ij})
	    ({\bm \sigma}_j \cdot \hat{\bm R}_{ij})
	    }{R_{ij}^5} \nonumber\\
	    &+ \sum_{i=1}{\bm \sigma_i}\cdot {\bm \nabla}\psi({\bm R}_i) ~.
\label{eq_Heff}
\end{align}
where $\psi({\bm R}_i)$ is the potential created by the virtual fully occupied system, where we dropped a background constant energy
term. This potential becomes constant for an infinite or periodic system of single coordination.

The approximate Hamiltonian ${\cal H}$ then describes the effective spin-spin interactions emerging from the dipolar repulsive force between the colloids, and it is analogous to the dipolar spin ice Hamiltonian, although it displays a spatial dependence of $R_{ij}^{-5}$ instead of $R_{ij}^{-3}$.
For both models, first-neighbor interactions are ferromagnetic (FM) (of the form $J_1{\bm \sigma}_i \cdot {\bm \sigma}_j$ with $J_1=-78Jd^2/a^5$ in the present case), and second-neighbor interactions are weakly antiferromagnetic ($J_2 = 14Jd^2/(3\sqrt{3} a^5) \approx 0.03|J_1|$).
It is known that first-neighbor FM interactions enforce the ice rule at low temperature, suppressing vertices with $Q=\pm 3$, and leading to the SI1 phase~\cite{willsetal2002}. 
Due to spin disorder, this phase has a finite entropy ($s\approx 0.501$ per spin~\cite{willsetal2002}). 
Weakly antiferromagnetic second-neighbor interactions induce a lower energy phase in which charged vertices are ordered in two staggered triangular sublattices of opposite charge~\cite{willsetal2002,Che11, canalsetal2016}.
This phase, called the SI2 phase, maximizes the distance between similar charges, but it is still spin disordered and has an extensive entropy of $s\approx 0.108$~\cite{Che11}.
Then, at an even lower critical temperature, long-range spin order emerges, giving rise to the chiral spin solid phase. Thus, it is natural to expect that the Hamiltonian expressed in Eq.~\eqref{eq_Heff} produces the same series of phases, with a similar ground state. 
In the following, we
indeed confirm this expectation by running simulations with
very tight traps. However, the low-temperature physics differs
for soft traps, stabilizing instead a FM ground state.

\textit{Monte-Carlo results.--}
We perform Monte-Carlo (MC) simulations in continuous space with periodic boundary conditions, modeling the double-well potential by a quartic function that describes two locally isotropic potential wells, Fig.~\ref{Figure1}(b) 
\footnote{We use the function 
$U(x, y) = h \left[\left(2x/d) \right)^2 - 1 \right]^2 + 16h \left(y/d \right)^2$. 
$x = {\bm \delta r}_\parallel$ and $y = {\bm \delta r}_\perp$ are the projections of the distance between the colloid and the center of the double-well onto the directions parallel and perpendicular to the axis of the double-well, respectively. 
The parameters of the quartic function are chosen such that the two minima are located at $(\pm d/2, 0)$, the height of the hill separating them is $U(0,0)=h$, and the potential is locally isotropic around the two minima.}.
On each step of the MC we realize three types of moves:
1) small particle moves sampled from an isotropic Gaussian distribution of width $\ll d$;
2) single spin flips, which consist in moving a particle to the symmetric position with respect to the center of the double-well; and
3) loop updates, where a loop of head-to-tail spins is simultaneously flipped.
Note that moves 2) and 3) leave the potential energy unchanged, and only modify the interaction energy. 
Also, we notice that loop updates are necessary to efficiently sample low-energy configurations of this frustrated system. 
In the limit $h/J \to \infty$ continuous motion of particles will be energetically forbidden and the colloids will remain in the minima of $U({\bm r_i})$. This approaches a discrete model in which the energy is given by the first term of Eq.~\eqref{H_coll}, namely, the interaction energy between particles.

\begin{figure}[t]
\begin{center}
\includegraphics[width=\columnwidth,keepaspectratio]{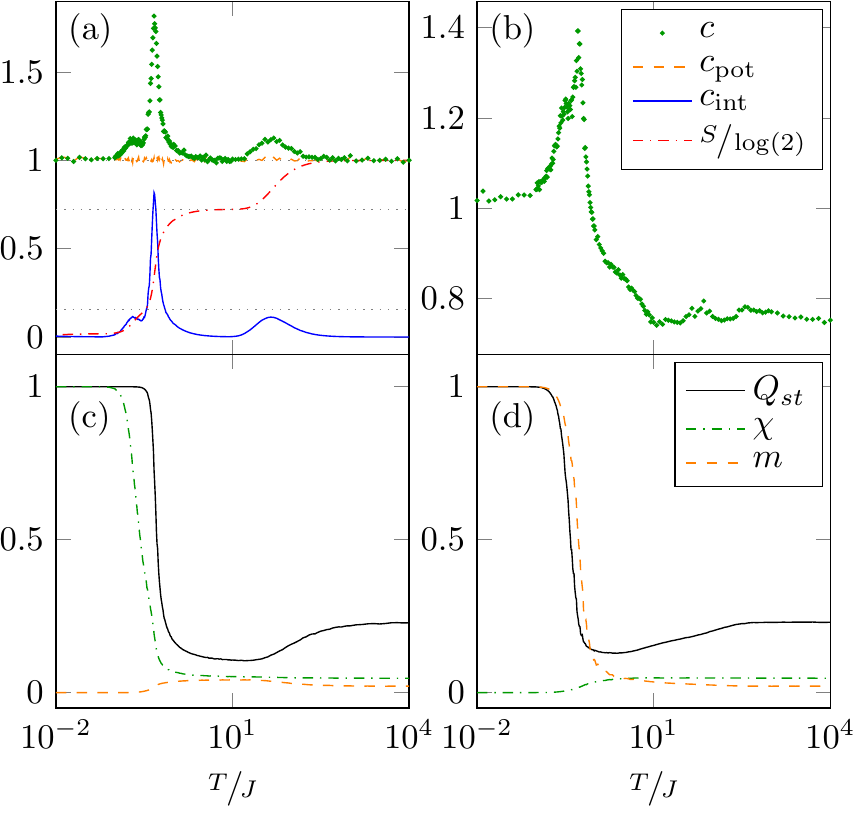}
\caption{Specific heat $c$ (a,b), and order parameters (c,d) resulting from Monte-Carlo simulations of the Hamiltonian ${\cal H}_{\rm coll.}$, for $d/a=0.3333$ and $N=3\times 6^2$ double-wells. (a,c): Discrete limit ($h/J=10^8$); (b,d) continuous limit ($h/J=10$). Panels (c, d) show the evolution of the staggered charge ($Q_{\rm st}$), chiral ($\chi$) and ferromagnetic ($m$) normalized order parameters (see text for their definitions).}
\label{Figure3}
\end{center}
\end{figure}
In Fig.~\ref{Figure3}(a,c), we present results of our MC simulations with $h/J=10^8$ and $d=0.3333a$, corresponding to the discrete limit.
The heat capacity is calculated from the fluctuations of the energy,  $c={\rm var}(E) / T^2$, and Fig.~\ref{Figure3}(a) shows that it contains a constant contribution from the harmonic traps $U({\bm r_i})$.
By equipartition, this contribution is $1/2$ per degree of freedom (two spatial directions per colloid in our case).
In the limit of very tight traps, such fluctuations are essentially decoupled from the spin configuration of the system, and we can decompose the heat capacity into a contribution from the potential, and a contribution from the interactions:
$c\approx c_{\rm pot}+ c_{\rm int}$,
with $c_{\rm pot (\rm int)}={\rm var}(E_{\rm pot (\rm int)}) / T^2$.
An effective entropy for the spin degrees of freedom may thus be extracted from the interaction component as $S(T) = S(T_0) + \int_{T_0}^{T_1}{\rm d}T c_{\rm int}/ T$. 
The evolution of $S$ is plotted on Fig.~\ref{Figure3}(a) (red dashed-dotted line), revealing the same phases as the dipolar spin ice model \cite{Che11,canalsetal2016}: 
upon decreasing the temperature, the paramagnetic phase ($S = \log 2$) crosses over to the ice-rule SI1 phase ($S\approx 0.501$ \cite{willsetal2002, Che11, canalsetal2016}),
then a large peak in the heat capacity at $T_c^{(1)} \approx 0.7J$ indicates the transition to the charge-ordered SI2 phase ($S\approx 0.108J$) \cite{Che11, canalsetal2016},
and finally, the small peak at $T_c^{(2)}\approx 0.3J$ indicates the emergence of long-range spin order in the ground state~\cite{willsetal2002}.

This scenario is further confirmed by the evolution of the staggered-charge order parameter (OP) $Q_{\rm st}$, and the chiral-phase OP $\chi$.
The staggered charge OP is $Q_{\rm st} = N_v^{-1}\sum_{v}(-1)^{P_v} Q_v$, where the summation is performed over all the vertices $v$. 
$N_v=2N/3$ is the total number of vertices.
Here $P_v=1$ if $v$ belongs to the $A$-sublattice of the hexagonal lattice, and $-1$ on the $B$-sublattice.  The evolution of $Q_{\rm st}$ in Fig.~\ref{Figure3}(c) indeed shows the emergence of charge order below $T_c^{(1)}$. 
The onset of long-range order (LRO) in the chiral phase may be quantified by the (normalized) magnetic structure factor at wave-vector ${\bm k}_\chi = (4\pi / 3, 0)$:
\[
	\chi = \frac{1}{\bar \chi}\frac{1}{N^2} \sum_{i, j=1}^N \langle {\bm \sigma}_i \cdot {\bm \sigma}_j \rangle e^{{\bm k}_\chi \cdot ({\bm r}_i - {\bm r}_j)} ~,
\]
where $\bar \chi = 0.19753\dots d^2$ is such that $\chi=1$ in the chiral phase.
The evolution of $\chi$ is plotted on Fig.~\ref{Figure3}(c) and shows the emergence of LRO below $T_c^{(2)}$. On the same Fig.~\ref{Figure3}(c), finally, we plot the (normalized) FM OP given by
\[
	m = \frac{1}{\bar m}\frac{1}{N^2} \sum_{i, j=1}^N \langle {\bm \sigma}_i \cdot {\bm \sigma}_j \rangle ~,
\]
where $\bar m = (2d/3)^2$ is such that $m=1$ in the FM state, and vanishes at low temperature. We observe the same qualitative behavior for any separation $d$ between the two minima of a double-well, for sufficiently large $h$.

Fig.~\ref{Figure3}(b,d) shows the results of MC simulations for $h/J=10$ and $d=0.3333a$.
In this regime, colloids can continuously move away from the bottom of the trap, and we cannot separate the fluctuations of $E_{\rm int}$ and $E_{\rm pot}$, as both are strongly (anti-)correlated and, as a consequence, we cannot reconstruct the entropy for the spin degrees of freedom from energy fluctuations.
Nevertheless, in Fig.~\ref{Figure3}(b), the heat capacity shows a clear peak at $T/J\approx 0.8J$, signaling the onset of LRO.
In Fig.~\ref{Figure3}(d) we observe that this LRO corresponds to the emergence of FM order, while the chiral-phase OP vanishes. Interestingly, in contrast to the discrete limit, we do not clearly observe an intermediate SI2 phase with charge order but no spin order. This suggests a direct transition from the SI1 phase to the FM phase, and confirming this hypothesis would require an extensive simulation  study beyond the scope of the current article. We note that FM order has also been observed in Brownian simulations of the same model \cite{libaletal2018}, although it was not identified as a consequence of the continuous degree of freedom of the colloids \cite{nisoli2018}. 

\begin{figure}
\centering
\includegraphics{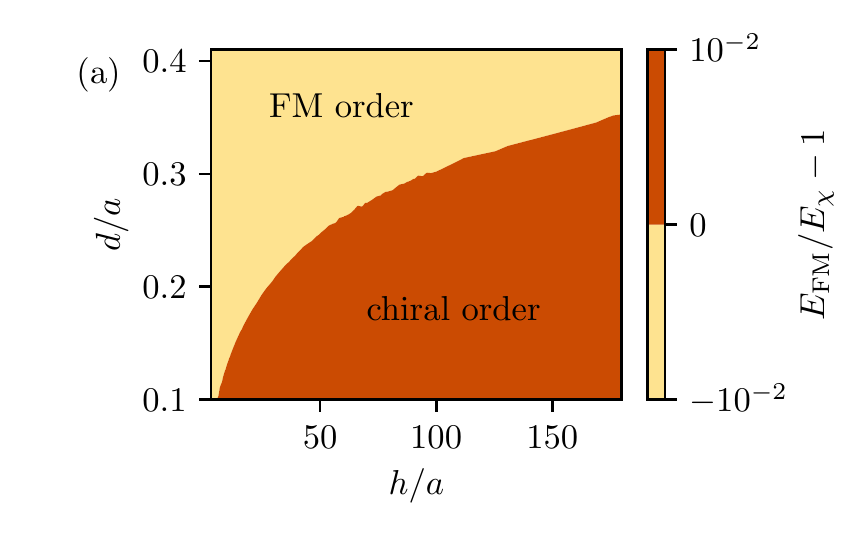}
\includegraphics{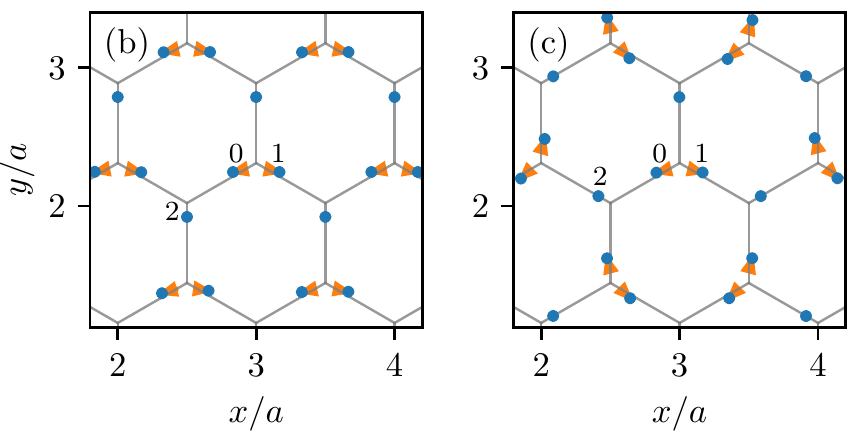}
\caption{
(a) Phase diagram of the colloidal ice in terms of the relative difference in energy between the ferromagnetic ($E_{\rm FM}$) and chiral ($E_\chi$) states.
(b,c) Particle displacements in the double wells ($d=0.4$, $h=20$)
when arranged initially in a ferromagnetic (b) and chiral (c) state. Dots denote particle positions, arrows direction of displacement. 
Colloids marked $1$ and $2$ are respectively first and second neighbors of colloid $0$ (see text).
}
\label{Figure4}
\end{figure}	

\textit{Ground-state phase diagram.--} Since the FM and chiral order break different discrete symmetries of the Hamiltonian, they are incompatible. 
At least one phase transition must occur in the ground state upon changing the shape of the double-well (namely the distance $d$ between the two minima, and the height $h$ of the hill between them).
In Fig.~\ref{Figure4}(a) we show the lowest energy configuration obtained for the different geometric parameters $h$ and $d$. 
These values are calculated by comparing the minimum energy of the FM state with that of the chiral order state, once the particles are allowed to relax in continuous space. 
For every point we preformed MC simulations starting from each configuration, with particles in the minimum of the double well, but allowing no spin flips, only continuous particle motion. 
We observe that FM order is always favored at smaller $h$, while for decreasing $d$ the chiral phase emerges.

The change in the ground-state configuration can be understood by considering the displacement of the colloids from the bottom of the traps as illustrated in Figs.~\ref{Figure4}(b,c). 
The more noticeable effect occurs at 2\textit{in} vertices, where the two incoming particles (here labeled as $0$ and $1$) move away from each other, while particles at 1\textit{in} vertices barely move.
In both cases, the particles in the 2\textit{in} vertices move away from each other, slightly deviated by the potential towards the axis of the double-well. The energy gain of the particles in the vertex and from the potential is similar in the case of the FM and the chiral order. 
The more important difference originates from the interactions with next nearest neighbor.
The off-axis component of the displacement leads to a more favorable interaction between particle $0$ and particle $2$ in the case of the FM configuration,
which leads to a reversal of the energy hierarchy among the two states. 
This explains why the FM phase is favored for larger values of $d$ (for a fixed $h$): 1st-neighbor interactions are larger for larger $d$, inducing a larger displacement of the colloids, activating the above-mentioned mechanism.

\textit{Conclusion.--} In summary, we have shown that for the kagome colloidal ice the motion of the colloids away from the bottom of the traps can modify the ground-state configuration, favoring a ferromagnetic instead of a chiral ground state. 
The shape of the double-wells thus has a fundamental influence on the hierarchy of low-energy states.  The novel phenomenology uncovered in our study naturally opens several directions for future research. First, one could further explore the nature of the phase transition to FM order at low temperature. Then, a natural question concerns the transition between ferromagnetic and chiral order upon changing the shape of the double wells. Finally, one may suspect that the motional degree of freedom of the colloids inside the traps can modify the low-energy physics of other frustrated models, \textit{e.g.} the triangular- and square-lattice spin-ice models. 

\begin{acknowledgments}
We thank Nicolas Rougemaille for many stimulating discussions. 
This work was supported by the European Research Council (Proj. No. 335040). P.T. acknowledges support from Mineco
(Proj. No. FIS2016-78507-C2-2-P) and AGAUR (Proj. No. 2017SGR1061) and Generalitat de Catalunya under Program "ICREA  Acad\`emia".
I.F. acknowledges support from Mineco (National Plan
15 Grant: FISICATEAMO No. FIS2016-79508-P, SEVERO OCHOA No. SEV-2015-0522, FPI), European Social Fund, Fundaci{\'o} Cellex, Generalitat de Catalunya (AGAUR Grant No. 2017 SGR 1341 and CERCA/Program), ERC AdG OSYRIS, EU FETPRO QUIC, and the National Science Centre, Poland-Symfonia Grant No. 2016/20/W/ST4/00314. 

\end{acknowledgments}
\bibliography{biblio}

\begin{thebibliography}{35}
\bibitem [{Val()}]{Values}%
  \BibitemOpen
  \href@noop {} {}\bibinfo {note} {We find for the ferromagnetic phase
  $\alpha=49.77804(2)$, $\beta=-122.85(1)$ and $\gamma=746.6(66)$ and for the
  chiral phase $\alpha=49.69919(2)$, $\beta=-124.49(2)$ and
  $\gamma=752.3(67)$}\BibitemShut {NoStop}%
\end{thebibliography}
\end{document}